# Effect of the FAST NUCLEAR ELECTROMAGNETIC PULSE on the Electric Power Grid Nationwide: A Different View

**Mario Rabinowitz**

*Electric Power Research Institute*

*Palo Alto, California 94303*

Inquiries to: *Armor Research*

*715 Lakemead Way, Redwood City, CA 94062*

Mario715@earthlink.net

**Abstract**

This paper primarily considers the potential effects of a single high-altitude nuclear burst on the U.S. power grid. A comparison is made between EMP and natural phenomena such as lightning. This paper concludes that EMP is no more harmful to the power grid than its counterparts in nature. An upper limit of the electric field of the very fast, high-amplitude EMP is derived from first principles. The resulting values are significantly lower than the commonly presented values. Additional calculations show that the ionization produced by a nuclear burst severely attenuates the EMP.

**INTRODUCTION**

The first of the last five U.S. atmospheric nuclear bomb tests took place in 1962, when a 1.4-megaton (Mt) hydrogen bomb detonated 397 km (248 mi) above the Johnston Island area. Since this test, known as Starfish, many reports have incorrectly stated that EMP produced a major electrical disturbance in Hawaii [1-5]. Most reports place the blast as 1,280 km (800 mi) from Hawaii, but this may also be incorrect.

← Out of more than 10,000 streetlights on the Hawaiian island of Oahu at the time of the Starfish burst, some 10 to 30 disc cutout fuses turned off as many streetlight strings of about 10 lights per string. Thus, only between 1% and 3% of the streetlighrs malfunctioned. The Hawaiian Electric Company reported no breakdown of power lines nor opening of circuit breakers, nor is there any indication from any source that such things occurred. The disc cutout fuses were easily restored and there was no damage to the streetlights or any streetlight or power circuits [6]. If the EMP caused the outage of - 1% of the streetlights of an already troublesome and antiquated system, then its potency was barely at the threshold of a system that failed regularly on its own.

← Similarly, communications systems such as telephone, radio, and television in Hawaii remained in full operation.

→ Starfish and the other four nuclear bomb bursts produced few if any EMP effects in Hawaii or even at Johnston Island much closer to the bursts.

→ A 1983 Sandia Laboratories report commissioned by the U.S. Nuclear Regulatory Commission analyzed the "worst case" scenario and concluded that EMP poses no substantial threat to nuclear power plants based upon both analysis and simulated EMP tests [7]. Since no real differences exist between electrical equipment in nuclear and other power plants, EMP effects are likely to be minimal in both cases.

## A QUESTION OF EFFECT

Of the two basic kinds of EMP, one is a relatively slow electro-magnetic pulse called magnetohydrodynamic EMP(MHD EMP), lasting $< \sim 10^2$ sec., $E < \sim 10^{-1}$ V/m. This is similar to solar storms which last $\sim 10$ min, $E > \sim 10^{-2}$ V/m. The other gives rise to a quick pulse referred to herein as "TEMP." Reference [8] describes the various kinds of EMP. It is important to distinguish the true effects of EMP from effects that arise from other sources.

### The Argus Effect

EMP is not the only electromagnetic effect associated with a high-altitude nuclear burst. One other occurrence is the formation of man-made radiation belts around the earth. These belts interfere with communications by affecting the medium through which radio waves propagate, and by direct interaction of the high-energy electrons in damaging components of satellites. The belts produced by a nuclear burst cannot affect the power grid, however, and are not part of what has been coined "the electromagnetic pulse."

### Artificial Effects

The Senate Hearings of 1963 [91], containing an extensive look at the Starfish blast and its effects in Hawaii, calls into question some claims of an EMP blackout in Hawaii. In particular, the report -- published one year after the burst -- does not mention the popping of circuit breakers, the downing of power lines, or any electrical disturbance on the Hawaiian utility system. Streetlights are not even mentioned. However, the report does discuss the tripping of burglar alarms in Honolulu, suggesting an alternate explanation for the occurrence (Ref. (9], p. 200).

## LIGHTNING VERSUS EMP

Recent measurements indicate that lightning strokes contain significant components with rise times $< 10^{-7}$ sec [10]. Lightning creates electric fields in the discharge region $> 10^6$ V/m, over an order of magnitude higher than the peak TEMP field, with power levels $\sim 10^{12}$ W, and energy dissipation between $10^9$ and $10^{10}$ J. Instruments an orbiting satellites have detected many lightning bolts with currents as high as $10^6$ A, which the above numbers do not even take into account.

A direct lightning strike produces the most severe effects. The high current density $\sim 10^3$ A/cm$^2$ in a stroke delivers a high power density to the strike point resulting in demolished structures such as exploded timber, molten metal, and charred insulation. Lightning transients can propagate along transmission lines at almost the speed of light, with circuit limited rise times - $10^{-6}$ see to peak voltages as high as - million volts, with a maximum rate of rise of ~ $10^{12}$ V/sec. In fact, a lightning strike on the 110-kV line of the Arkansas Power & Light Company [11] reached a peak voltage of 5 million volts within 2 µsec without calamitous results.

The spectacular destruction accompanying a direct lightning hit is not apt to accompany EMP. The primary effect of the TEMP is to induce overvoltages and overcurrents in the power system. The transients produced in the system will have a slower rise time than the free wave TEMP. The newly developed zinc oxide lightning arrestors should be quite effective in shunting these pulses to ground. It has been shown in Ref. [81 that these induced voltages and currents cannot cause flashover across transmission lines, and can only do so in isolated instances for distribution lines. The TEMP excites all three phases and the ground wire(s) similarly and almost simultaneously. Thus, it cannot produce large enough voltage differences between the lines to produce breakdown. This consideration, together with those of leaders and streamers, electron time of flight, and electron avalanche [8] imply that it is impossible for the EMP to cause widespread damage to the transmission line system, and almost as likely for the distribution grid.

## HOW MUCH ENERGY CAN THE TEMP HAVE?

The TEMP energy cannot exceed the total energy of the prompt gamma rays since they give birth to it by creating the Compton recoil electrons. Conventional analyses give the fraction of energy accounted for by prompt gamma rays as between 3 x $10^{-3}$ [121 and $10^{-4}$ [13]. Thus, $10^{-3}$ is a representative number. The electric field strength, E, in the pulse as a function of time, t, is generally given as [7]:

$$E = E_o (e^{-\alpha t} - e^{-\beta t}), \qquad (1)$$

where the parameters in this Design TEMP are

$$E_o = 52.5 \times 10^3 \text{ V/m},$$
$$\alpha = 4.0 \times 10^6 /\text{sec},$$
$$\beta = 4.78 \times 10^8 /\text{sec}.$$

From the time integral of the Poynting vector, $\int_0^t \vec{E} \times \vec{H} \, dt$, one can calculate the total energy density, D, delivered by the TEMP during the time, t.

$$D(t) = \int_0^t \frac{E_o^2}{Z} (e^{-\alpha t} - e^{-\beta t})^2 dt, \quad Z = 377 \, \Omega$$

Thus, $D(10 \text{ nsec}) = 4.8 \times 10^{-2} \text{ J/m}^2$

$$D(1 \text{ µsec}) = 8.9 \times 10^{-1} \text{ J/m}^2. \quad (2)$$

In doing an energy balance, let us look at a 1.4-Mt bomb in precisely the Johnston Island scenario here and throughout this report since larger bombs do not produce greater magnitude TEMP [7,12,14]. The total energy release of this bomb is $5.9 \times 10^{15}$ J. Therefore, the total gamma ($\Gamma$) energy is:

$$\Sigma_{\Gamma total} = 10^{-3}(5.9 \times 10^{15} \text{ J}) = 5.9 \times 10^{12} \text{ J} \quad (3)$$

With the energy per unit area delivered by the TEMP, as calculated in Eq. 2, It is now possible to calculate the total energy delivered. Not only have the media treated the TEMP as having a magnitude of 50 kV/m from border-to-border and from coast-to-coast, but this scenario has also been promulgated at technical meetings and in technical papers and reports. One technical paper[15] makes the following statement: "The early time HEMP [TEMP] produced by a high altitude nuclear detonation behaves as a plane electromagnetic wave that sweeps the earth's surface." The paper goes on to use the "Design Pulse" described by Eq. 1: "A representative double exponential EMP electric field waveform given by Eq. 1 was used to calculate representative HEMP induced surges."

The calculations in this report neither support the magnitude of the Design Pulse, nor its ubiquity as a "plane electromagnetic wave that sweeps the earth's surface." Nevertheless, by calculating explicitly the implicit energy content in this prescription, we will see that it leads to a violation of the conservation of energy by over two orders of magnitude.

Using the plane wave prescription which is commonly (incorrectly) assumed, the energy delivered by the TEMP at the end of 1 µsec is:

$$\Sigma_{TEMP} = 8.9 \times 10^{-1} \text{ J/m}^2 (1.4 \times 10^{13} \text{ m}^2) \quad (4)$$
$$= 1.2 \times 10^{13} \text{ J}.$$

$$\Sigma_\Gamma = 0.34 (5.9 \times 10^{12} \text{ J}) = 2.0 \times 10^{12} \text{ J}. \quad (5)$$

$$\frac{\Sigma_{TEMP}}{\Sigma_\Gamma} = \frac{1.2 \times 10^{13} \text{ J}}{2.0 \times 10^{12} \text{ J}} = 6 = 600\% \quad (6)$$

It is impossible for the energy conversion process from gamma rays to Compton electrons to TEMP to be 600% efficient, or even 100% efficient. Even taking into account minor spatial variations of the TEMP (Ref.[121, p. 538) reduces the discrepancy only by a factor of 2.5, but a factor of 600 (60,000%) must be accounted for. According to Ref. [121, p. 534, there is a 1% energy conversion efficiency of the prompt gamma rays to EMP. Thus, the analysis in this section shows that the energy content Implicit In the "Design Pulse" is too high by a factor of 600.

**POWER RADIATED PER ELECTRON**

The power radiated by a nonrelativistic accelerated electron is

$$P = \frac{e^2 a^2}{6 \pi \varepsilon_o c^3}, \qquad (7)$$

where a is the acceleration and $\varepsilon_o$ is the permittivity of free space. Since the electrons under discussion travel near the speed of light, the radiation is enhanced by relativistic effects, and the radiated power is

$$P = \frac{e^2}{6 \pi \varepsilon_o c^3} \gamma^4 \{a^2 + \left(\frac{\gamma}{c}\right)^2 (\vec{v} \cdot \vec{a})^2\}. \qquad (8)$$

During the interval of maximum power radiationj most of the electrons have circular or helical orbits of roughly constant pitch and radius, where v is perpendicular to a, so $\vec{v} \cdot \vec{a} = 0$. Before losing much energy, the I-MeV electrons if acting independently will each be radiating at most about $4 \times 10^{-22}$ Watts.

It Is interesting to compare the average power radiated per electron as derived from the given TEMP characteristic with the relativistic calculation of the maximum synchrotron radiation from one electron.

The power density in the TEMP is

$$P/A = \vec{E} \times \vec{H} = \frac{E_o^2}{Z} (e^{-\alpha t} - e^{-\beta t})^2. \qquad (9)$$

The maximum power density occurs at about 10 nsec, with a value of 6.6 MW/m$^2$ = 6.6 × 10$^6$ W/m$^2$.

Since the prompt gamma rays show spherical symmetry about the burst, let us consider a very narrow cone whose vertex is centered at the burst and whose base is 400 km (250 mi) from the burst, and has an area of 1 m$^2$. The power radiated by the TEMP over this 1 m$^2$ area at 10 nsec is

$$6.6 \times 10^6 \text{ W/m}^2 \text{ (1 m}^2\text{)} = 6.6 \times 10^6 \text{ W.} \qquad (10)$$

The maximum number of Compton electrons of 1-MeV average energy in this cone is

$$N \sim \frac{\Sigma_{\Gamma_{total}}/4\pi R^2}{\text{energy/electron}} = \frac{5.9 \times 10^{12} \text{J}/4\pi(400 \times 10^3 \text{m})^2}{10^6 \text{eV/electron}}$$

$$= 2 \times 10^{13} \text{ electrons.} \qquad (11)$$

The TEMP power per electron is

$$\frac{6.6 \times 10^6 \text{ W}}{2 \times 10^{13} \text{ electrons}} = 3 \times 10^{-7} \text{ W/electron.} \qquad (12)$$

$$\frac{\text{TEMP Radiation}}{\text{Synchrotron Radiation}} = \frac{3 \times 10^{-7} \text{ W/electron}}{4 \times 10^{-22} \text{ W/electron}}$$

$$= 8 \times 10^{14}.$$

This number implies a degree of coherence (laserlike effect) between the radiating electrons that is highly unlikely In an uncontrolled environment.

## REASONABLY FUNDAMENTAL LIMITATION TO EMP

At any instant the energy radiated by the ensemble of Compton electrons is stored equally in the electric and magnetic fields in the oscillating electromagnetic waves. This energy cannot exceed the accumulated earth-directed part of the gamma energy at any instant of time. Thus

$$\int \frac{1}{2} \epsilon_o E^2 dV + \int \frac{1}{2} \frac{B^2}{\mu_o} dV$$

$$= 2 \int \frac{1}{2} \epsilon_o E^2 dV = f \Sigma_{\Gamma e}(t), \qquad (13)$$

where f is the average conversion efficiency from gamma rays to EMP.

The volume element dV = Acdt, and taking the time derivative of Eq. 13 gives:

$$\epsilon_o E^2 Ac = f \frac{d}{dt} \Sigma_{\Gamma e} = f P_{\Gamma e}. \tag{14}$$

$P_{\Gamma e}$ is the power in the prompt gamma flux toward the earth, and $P_{\Gamma e} = \frac{1}{3} P_{\Gamma}$.

From Ref. [12], p. 328, one can estimate that the maximum gamma power from a 1.4-Mt bomb is

$$P_{\Gamma} = 10^{20} \text{J/sec} = 10^{20} \text{ Watts}. \tag{15}$$

Therefore combining Eqs. 14 and 15:

$$E_{max} = \left[ \frac{f \times 10^{20} \text{J/sec}}{3\epsilon_o Ac} \right]^{1/2}. \tag{16}$$

Only ~ $10^{-2}$ of the prompt gamma energy is converted to TEMP [12, p. 534], and A = 1.4 x $10^{13}$ m$^2$ (corresponding to a burst height of 400 km) in Eq. 17 yields

$$E_{max} = 3 \text{ kV/m}. \tag{17}$$

Eq. 17 gives the average value of the maximum TEMP electric field over an area comparable to that of the United States.

It is also possible to calculate an upper limit of the electric field at any point, since the maximum prompt gamma flux exceeds the Poynting vector of the TEMP at every point:

$$|\vec{E} \times \vec{H}|_{max} = \frac{E_{max}^2}{z} = \frac{f P_{\Gamma}}{4\pi r^2}, \tag{18}$$

z = 377 Ω. Eq. 18 implies

$$E_{max} = \left[ \frac{f P_{\Gamma} z}{4\pi} \right]^{1/2} \frac{1}{r} = \left[ \frac{f z \ 10^{20} \text{J/sec}}{4\pi} \right]^{1/2} \frac{1}{r} \tag{19}$$

at any point on the earth's surface, which is a distance r from the burst. From Eq. 19, Table I presents $E_{max}$ as a function of r for the given f = $10^{-2}$ and for a factor of 10 increase in the product of gamma power and conversion efficiency in the second column.

TABLE I

$E_{max}$ at Various Distances from a 400-km
High 1.4-Mt Burst

| r km | f = 1%<br>$E_{max}$ kV/m | 10 f$P_\Gamma$, 10 $P_\Gamma$<br>to f = 10%<br>$E_{max}$ kV/m |
|---|---|---|
| 900 | 5.1 | 16 |
| 1200 | 3.8 | 12 |
| 1440 | 3.2 | 10 |
| 1600 | 2.9 | 9.2 |
| 1800 | 2.5 | 7.9 |
| 2000 | 2.3 | 7.3 |
| 2200 | 2.1 | 6.6 |

Figure 1 compares the Design Pulse with the spatial average value of E max of 12 kv/m for the Column 2 case, which allows an extra factor of 10 in f$P_\Gamma$. Clearly, the EMP is still significantly smaller than the Design Pulse even for this enhanced case.

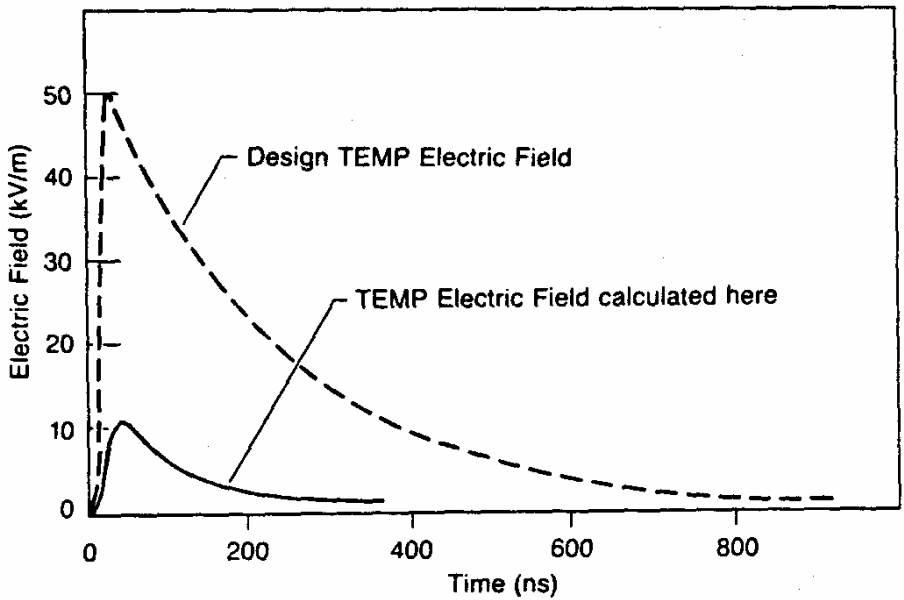

Figure 1: TEMP Electric Field

Using the EMP from this factor of 10 bigger or more efficient bomb, Figure 2 compares the voltage induced on a transmission line by the pulse calculated in this report and the Design Pulse with lightning.

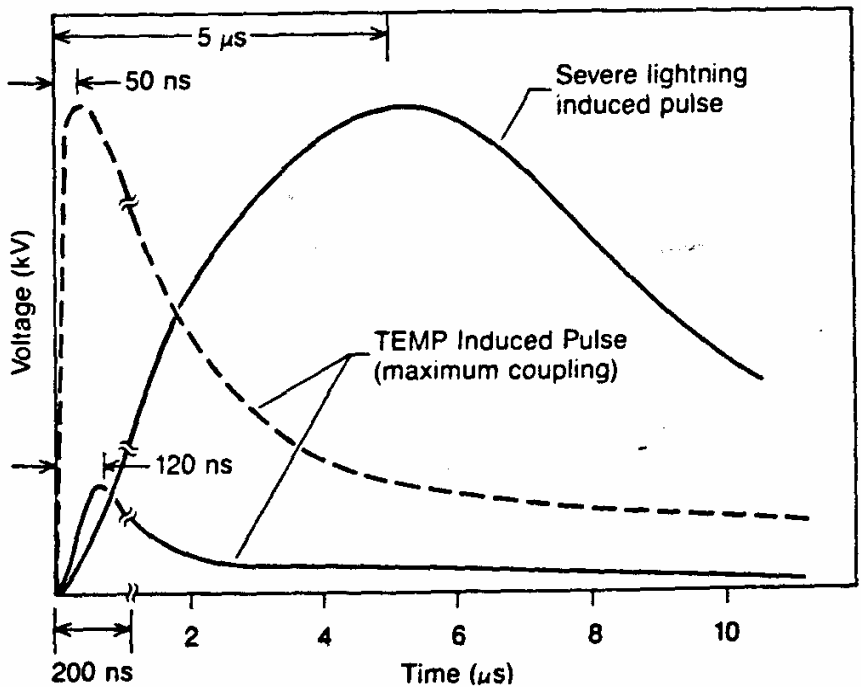

Figure 2: TEMP Induced Pulse (maximum coupling)

Figure 2 shows that the TEMP-induced pulse is lower than a severe lightning-induced pulse over most of the timeframe that is crucial for breakdown.

## TEMP ATTENUATION AND PHASE SHIFTS DUE TO ATMOSPHERIC CONDUCTIVITY

Atmospheric Conductivity

The source region for the production of TEMP from a high-altitude burst, to a good approximation, occurs primarily in a 10 km-thick region of the atmosphere roughly between 25-km and 35-km altitude. Well above a 35-km height, the density is too low for much production of Comptons. Below 25 km, most of the prompt gammas have been absorbed, and again there is not much production of Comptons. In the source region, there is appreciable ionization by the Comptons in producing secondary electron-ion pairs as discussed by Longmire [16]. He says: "Each of the Compton recoil electrons, which have starting energies of the order of I MeV, generates of the order of 30,000 secondary electron-ion pairs along its track in the air" (Ref. [16], p. 3). There will also be considerable ionization produced by the X-rays from the bomb. This effect will not be included at this time. The following calculations help us determine whether the source region can be treated as a good conductor, and also estimate its conductivity.

There are two conditions appropriate to a good conductor. These are:

1. Collision frequency is much greater than the-oscillation frequency: $\tau^{-1} \gg \omega$.

2. Conduction current is much greater than displacement current: $\sigma E \gg \omega \varepsilon E$, i.e., $\sigma \gg \omega \varepsilon$.

Does the TEMP source region meet these two conditions?

From the Standard Atmosphere [17], we find that the mean collision time, $\tau_M$, between molecules at an altitude of 30 km is $10^{-8}$ sec. In general, the mean collision time is $v = (nCv)^{-1}$ where n is the number density of scatters, C is the scattering cross section, and v is the mean velocity. Again, from Ref. [17], we find that the mean molecular velocity, $v_M$, at 30 km is 400 m/sec. Thus,

$$\tau = 10^{-8} \text{ sec } \left[\frac{C_M \, v_M}{C_e \, v_e}\right]. \tag{20}$$

For thermalized electrons, Eq. 20 becomes

$$\tau^{-1} = 10^8/\text{sec} \left[\frac{C_e}{C_M} \left(\frac{M}{m}\right)^{1/2}\right] \tag{21}$$

$$= 2 \times 10^{10} \left(\frac{C_e}{C_M}\right) \sim 10^9/\text{sec}.$$

$$\tau^{-1} \lesssim 10^{10}/\text{sec for unthermalized electrons.}$$

Now to estimate the highest frequency in the TEMP pulse. The TEMP reaches its peak value in about 10 nsec. To a good approximation, this is 1/4 period of the highest harmonic:

$$1/(4f) \sim 10^{-8} \text{ sec} \tag{22}$$

Hence, $f = 2.5 \times 10^7$ Hz. So

$$\tau^{-1} \sim (10^9 - 10^{10})/\text{sec} \gg \omega = 2\pi (2.5 \times 10^7 \text{ Hz})$$

$$= 1.6 \times 10^8/\text{sec}. \tag{23}$$

Therefore the first condition is met.

It is necessary to know the conductivity, σ, of the source region to determine if the second condition is met. Longmire [16] calculates that saturation occurs for the TEMP when:

$$\sigma > \frac{1}{2\pi h} \text{ cm}^{-1} \sim 10^{-6} \text{ mho/m.} \tag{24}$$

Now consider the second condition in the case of the highest frequency:

$$\omega\varepsilon \doteq \omega\varepsilon_o = 1.6 \times 10^8/\text{sec } (8.85 \times 10^{-12} \text{ farad/m}) \tag{25}$$

$$= 1.42 \times 10^{-3} \text{ mho/m.}$$

Excluding sources of ionization other than the prompt gamma rays, Condition 2 may not be met as well as Condition 1 for the highest frequencies, but may be adequately met for the lower frequencies. Since several sources give the number density of secondary electrons as a function of time, Eq. 26 would allow determination of the conductivity as a function of time,

$$\sigma = n_s e^2 \tau/m. \tag{26}$$

From Eq. 21, a conservative estimate is that $t > 10^{-10}$ sec at an altitude of 30 km. Using the values for the number density of secondary electrons, $n_s$, in Longmire ([16], Fig. 2), one can construct the following table for conductivity in the source region as a function of time.

TABLE II

Electrical Conductivity in Source Region

| t, nsec | $n_s$, $10^{12}$ elec/m$^3$ | σ, $10^{-6}$ mho/m |
|---------|------------------------------|---------------------|
| 2       | 0.2                          | 1                   |
| 5       | 1.0                          | 3                   |
| 10      | 2.0                          | 6                   |
| 25      | 3.5                          | 10                  |
| 50      | 5.5                          | 16                  |
| 150     | 10.0                         | 30                  |

Table II indicates that the TEMP saturates in 2 nsec. Using Longmire's own criteria [16] of saturation when σ > $10^{-6}$ mho/m, this means that the peak of the pulse occurs before 2 nsec; and that the pulse cannot exceed 30 kV/m from 2 nsec on. This finding is in contradiction to the Design Pulse, which is assumed to peak at 50 kV/m at 10 nsec.

Skin Depth, Phase Shifts, and Attenuation in the Source Region

In terms of the skin depth, $\delta = (\pi f \mu \sigma)^{-1/2}$ the electric field of a plane wave in the x-direction is

$$E_z = E_o e^{-(\frac{1+i}{\delta})x} e^{i\omega t} = E_o e^{-x/\delta} e^{i(\omega t - x/\delta)}. \qquad (27)$$

The electric field radiated downward by each group of Compton electrons enters a conducting region (created by the generation of secondary electrons) immediately below it. This is its original surface value $E_o e^{i\omega t}$. There is a tremendous attenuation and a significant phase shift as the wave travels through the source region. Table III indicates the skin depth, $\delta$, attenuation of the field, $e^{-x/\delta}$, and phase shift, $x/\delta$ radians, for various conductivities and frequencies for a distance $x = 500$ m. Blank spaces are left in the table where Condition 2 is clearly not met, i.e., the conduction current is not significantly larger than the displacement current under the restriction that no other sources of ionization other than prompt gamma rays are included. Table III shows that there are significant phase shifts in a distance of 500 m.

When Displacement Current Exceeds Conduction Current in the Source Region

Let us determine the attenuation of the TEMP in the source region when the displacement current exceeds the conduction current, i.e., when $\omega \varepsilon \gg \sigma$. From Newton's second law, the equation of the average motion of one secondary electron of charge $-q$ and mass $m$ in the presence of an oscillating electric field, $E = E_o e^{i\omega t}$, is

$$m \frac{d^2 x}{dt^2} = -m\nu \frac{dX}{dt} - qE_o e^{i\omega t}, \qquad (28)$$

where the second term represents the frictional or viscous force due to collisions, and $\nu$ is the collision frequency.

The steady-state solution of Eq. 28 is

$$X = \frac{qE_o}{m} \frac{e^{i\omega t}}{\omega(\omega - i\nu)}. \qquad (29)$$

The current density associated with the motion of n secondary electrons per unit volume in the source region is

$$J = n(-q) \frac{dX}{dt} = \frac{-nq^2}{m} \frac{iE}{(\omega - i\nu)} = \sigma E, \qquad (30)$$

where $\sigma$ is the conductivity of the source region. Therefore the conductivity in this situation is

$$\sigma = \frac{-inq^2}{m(\omega - i\nu)}, \qquad (31)$$

where the conductivity here is a complex number that reduces to a real number when $\omega \ll \nu$. For

$$\omega \ll \nu: \quad \sigma = \frac{-inq^2}{m\nu(\frac{\omega}{\nu} - i)} \doteq \frac{-inq^2}{m\nu(-i)} = \frac{nq^2}{m\nu} = \frac{nq^2 \tau}{m}, \qquad (32)$$

where $\tau$ is the mean time between collisions.

The propagation and attenuation constants, $\alpha$ and $\beta$, are found from the condition

$$T^2 \equiv (\alpha + i\beta)^2 = \mu\varepsilon\omega^2 + i\sigma\mu\omega$$

$$= \frac{\omega^2}{c^2}\left(1 + i\frac{\sigma}{\varepsilon_o \omega}\right). \qquad (33)$$

Under the condition that $\frac{\sigma}{\varepsilon_o \omega} \ll 1$, Eq. 33 implies that $\alpha \doteq \frac{\omega}{c}$, and the attenuation constant

$$\beta \doteq \frac{-n_s q^2}{2mc\varepsilon_o}\left(\frac{\nu}{\omega^2 + \nu^2}\right). \qquad (34)$$

Substituting numerical values into Eq. 34, and converting $\beta$ into an attenuation $a$ of db/km,

$$a = 4.6 \times 10^4 \left(\frac{n_s \nu}{\omega^2 + \nu^2}\right) \text{ db/km}, \qquad (35)$$

where $n_s$ is the number of secondary electrons/cm$^3$; $\nu$ is the number of collisions per second which an electron makes with ions, molecules, or atoms; and $\omega$ is the angular frequency of the radiated wave. For the source region at a height of 30 km, $\nu > \sim 10^9$/sec. Table IV gives the attenuation in decibels/km for the secondary electron number densities given by Longmire (Ref. [16], Fig. 2). . The same densities were used in Table II in per m$^3$. Entries are not omitted where the condition $\omega\varepsilon \gg \sigma$ is not met. This can be done on a case-by-case basis.

Even at early times, the attenuation for angular frequencies below $10^{10}$ Hz is significant. For example, at 5 nsec and $\omega = 10^9$ Hz, the attenuation is 23 db/km. To get the total attenuation rigorously, one should take an integral with respect to distance of the attenuation as a function of distance, taking into consideration the increase in collision frequency with decreasing altitude as well as the altitude dependence of the electron number density. However, to a fair approximation, simply multiply the attenuation at 30 km as given by Eq. 35 times the appropriate thickness depending on the angle the radiation goes through the source region. The radiation directed straight down from the Compton electrons at the top of the source region must travel 10 km through the source region. Its attenuation will be

$$(23 \text{ db/km}) (10 \text{ km}) = 230 \text{ db} = 10^{-23}. \tag{36}$$

There will be essentially nothing left of this radiation when it emerges from the source region.

## MULTIPLE BOMB EMP: PREIONIZATION AND POSTIONIZATION

Preionization

One may expect that the radiation from the fission bomb that triggers the fusion bomb will not produce much TEMP. However, it should produce preionization, i.e., secondary electrons in the source region that scale directly with its yield. It is unclear what size fission bomb would be used to trigger a fusion bomb. A reasonable guess is that it would be between 1 and 10 kilotons for a 1.4-Mt fusion bomb, with a prompt gamma yield ~ $10^{-2}$ kt.

To determine the effect of this small yield, one may scale the values given by Longmire ([16], Fig. 2). One may expect an overlap of the prompt gamma pulse from the fission trigger and the prompt gamma pulse during the fusion reaction, but this consideration will be neglected here. It is assumed that the prompt gamma pulse from the fission trigger bomb occurs at a prior time of between 2 nsec and 150 nsec. Entries have not been omitted where $\omega\varepsilon >> \sigma$ is not met. The attenuation in Table V is calculated from Eq. 35.

The number density of secondary electrons is given at a height of 30 km, assuming a geomagnetic field of 0.6 G, and 1.6 MeV prompt gammas. The attenuation is not negligible, although it is not nearly as great as in Table IV resulting from the total prompt gamma yield from a fusion bomb.

Postionization

Because the prompt gamma power density scales inversely with the square of the height of the burst above the source region, it may appear that many bombs will produce much more TEMP than a simple linear sum. For example, one might expect more than 50 times as much TEMP from 50 low-altitude bombs covering the United States than from one similar high-altitude bomb that covered the same area. Table V for h = 100 km indicates one reason why this is complicated and mollified by preionization.

TABLE V

Preionization Attenuation of the TEMP

| t, nsec | Burst at h = 100 km | | | Burst at h = 400 km | | |
|---|---|---|---|---|---|---|
| | $n_e$ $10^5$ e/cm$^3$ | $\omega=10^9$ Hz db/km | $\omega=10^8$ Hz db/km | $n_e$ $10^3$ e/cm$^3$ | $\omega=10^9$ Hz db/km | $\omega=10^8$ Hz db/km |
| 2 | 0.13 | 0.3 | 0.6 | 0.44 | 0.01 | 0.02 |
| 5 | 0.64 | 1.5 | 2.9 | 2.22 | 0.05 | 0.10 |
| 10 | 1.28 | 2.9 | 5.9 | 4.44 | 0.10 | 0.20 |
| 25 | 2.24 | 5.2 | 10.3 | 7.78 | 0.18 | 0.36 |
| 50 | 3.53 | 8.1 | 16.2 | 12.2 | 0.28 | 0.56 |
| 150 | 6.41 | 14.7 | 29.4 | 22.2 | 0.51 | 1.02 |

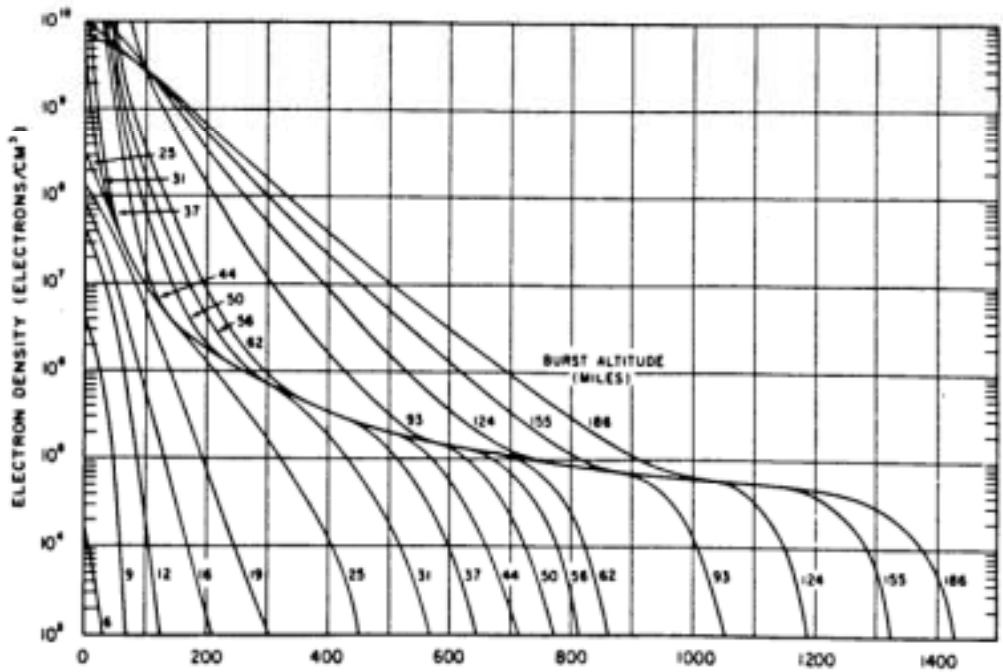

**HORIZONTAL DISTANCE (MILES)**
**Initial electron density at 40-miles altitude produced by prompt radiation from a 1-megaton explosion, as a function of distance (miles), for various burst altitudes.**

Figure 3: Electron Density from 1-Megaton Burst
Source: S. Glasstone and P. J. Dolan, The Effects of Nuclear Weapons, Washington, DC, U.S. Department of Defense, 1977, p. 499

The postionization of some 30,000 secondary electrons per primary Compton electron from one bomb plus all other sources of ionization sets the preionizition stage for a second bomb.

The conductivities, electron number densities, and attenuations were calculated in the previous sections for very short times and only the prompt gamma source of ionization.

As such they are orders of nagnitade smaller than the long-time values, which ionization is primarily produced by X-rays. In a report by Crain [18], which includes only γ-ray sources, we find in his Figure 2 that after a long time the number density of secondary electrons peaks at an altitude of 22 km at $6 \times 10^6$ secondary electrons/cm$^3$ with a value of $4 \times 10^6$ electrons/cm$^3$ at an altitude of 30 km, and $2 \times 10^5$ electrons/cm$^3$ at 64 km. This is for a burst height of 100 km, and a prompt gamma yield of $10^{-2}$ kt.

Figure 3, from Ref. [121, illustrates the wide ranging ionization mostly produced by x-rays from a 1-Mt explosion for a range of altitudes from 6 to 186 miles covering a horizontal distance from 0 to 1430 miles. The electron density is given at an altitude of 40 miles = 64 km, which is about double the altitude of interest. Scaling analysis indicates that these electron densities are approximately applicable to the source region.

Table VI gives a conservative estimate of the electron number densities and attenuation in the source region after a 1-Mt burst as a function of the horizontal distance R from the burst. The attenuation is calculated from Eq. 35. Entries have not been omitted where the condition $\omega\varepsilon >> \sigma$ is not met. When $\omega\varepsilon << \sigma$ the attenuation can be calculated from Eq. 27.

TABLE VI

Source Region Attenuation After 1-Mt, 186-Mi Altitude Burst

| R, miles | n, elec/cm$^3$ | $\omega = 10^{10}$ Hz db/km | $\omega = 10^9$ Hz db/km | $\omega = 10^8$ Hz db/km |
|---|---|---|---|---|
| 0 | $1 \times 10^{10}$ | $4.6 \times 10^3$ | $2.3 \times 10^5$ | $4.6 \times 10^5$ |
| 100 | $3 \times 10^9$ | $1.4 \times 10^3$ | $6.9 \times 10^4$ | $1.4 \times 10^5$ |
| 200 | $7 \times 10^8$ | $3.3 \times 10^2$ | $1.6 \times 10^4$ | $3.2 \times 10^4$ |
| 400 | $4 \times 10^7$ | 19 | $9.2 \times 10^2$ | $1.8 \times 10^3$ |
| 800 | $3 \times 10^5$ | 0.14 | 6.9 | 14 |
| 1300 | $3 \times 10^4$ | 0.014 | 0.69 | 1.4 |

**SENATE HEARINGS**

As mentioned earlier, the Senate Hearings of 1963 [9] shed considerable light on what happened during Starfish and the other nuclear bomb tests near Hawaii in 1962. Nowhere in the 981 pages of sworn testimony and Senatorial comments does it mention the popping of circuit breakers, the downing of power lines, or any electrical disturbance on the Hawaiian utility system. Streetlights are not even mentioned.

Among those testifying were Major General Robert H. Booth, Chief of the Defense Atomic Support Agency and Colonel Roy J. Clinton, acting Deputy Chief of Staff for Weapons Effects and Tests, who served as Booth's technical expert. The Defense Atomic Support Agency

(DASA), and the organizations from which the Agency is an outgrowth, have been connected with nuclear testing from the beginning ([9]. pp.157, 158).

Since these gentlemen headed DASA, the organization responsible for Starfish and the other Pacific nuclear tests, they are eminently qualified to testify on these tests. This is what they had to say (Ref. [9], pp. 164, 170, 171, 172):

> Senator Stennis...."You say that was a shot fired 400 km up in the air, and 800 km from Honolulu. Nevertheless, it set off burglar alarms in Honolulu.
>
> Colonel Clinton. Yes, sir.
>
> Senator Stennis. A kilometer is about three-fifths of a mile.
>
> General Booth. It was 500 miles from Honolulu, sir, and 250 miles in the air.
>
> Colonel Clinton. I should say, sir, this is a true statement. But burglar alarms also are very sensitive.
>
> Colonel Clinton. The Fish Bowl high altitude series was an enormously complicated series of events. Scientifically speaking, it was also one of the most successful.
>
> The concept for Star Fish was to launch, on a near-vertical trajectory, a Thor missile containing the device and three instrumentation pods.
>
> The pods were released on the way up so as to be positioned [deleted] from the burst; the missile reached an apogee of about 1,000 km, and the device was detonated on the way down at 400 km [deleted]."

Dr. Teller's paper [41, places the burst at, 600 miles from Hawaii. The Senate testimony confirms the distance of the burst as 500 miles from Honolulu is' probably more accurate than the commonly used figure of 800 miles.

**OBSERVATIONS**

Factors that contribute to a decrease in coherence are a large degree of scattering of the Compton electrons in nonradial directions by the gammas and atmospheric molecules; and frequency spread of the electrons due to a spread in their energies. Sollfrey [19] has calculated the effects of scattering on EMP. Other factors, such as the radiation reaction force and relativistic frequency shifts, will also contribute to a decrease in the magnitude of the TEMP.

The radiation reaction force, $F_r$, can be obtained from a simple consideration. In Eq. 12, it was found that the power radiated per electron is at least $3 \times 10^{-7}$ W/electron, for $E = 50$ kV/m.

$$\vec{F}_r \cdot \vec{v} = 3 \times 10^{-7} \text{ W.} \qquad (37)$$

With v - 0.9 c for a 1 MeV electron, Eq. 37 says that $F_r$ is at least $1.1 \times 10^{-15}$ N. The Lorentz force producing the radiation is $F_L = e\, \mathbf{v} \times \mathbf{B} = 2.6 \times 10^{-15}$ N. Since $F_r$ is almost as large as $F_L$, $F_r$ can significantly reduce the radiation [20].

In a large collection of radiating particles, there may be a number, N, of constructively coherent radiators. The Poynting vector of this coherenr ensemble is:

$$\left(\sum_{n=1}^{N} \vec{E}_n\right) \times \left(\sum_{n=1}^{N} \vec{H}_n\right) = (N\vec{E}) \times (N\vec{H}) = N^2 (\vec{E} \times \vec{H}), \quad (38)$$

where **E** and **H** are the ensemble averages. The total power is

$$P_{total} = \int \left(\sum_{n=1}^{N} \vec{E}_n\right) \times \left(\sum_{n=1}^{N} \vec{H}_n\right) dA$$

$$= N^2 \int \vec{E} \times \vec{H}\, dA = N^2 P, \quad (39)$$

where P is the typical power a single particle would radiate individually.

In the case of EMP, the constructive interference is essentially along a line-of-sight between the burst and the observer. Therefore, the average power radiated per electron is approximately $P_{total}/N = NP$. N times as much as power is radiated per electron, as a single independent electron can radiate.

This implies that the radiation reaction force, $F_r$, is N times greater than for an individual electron. Thus, the approximate relativistic expression for the radiation reaction force Is

$$\vec{F}_r = Nm_o \tau \gamma^4 \left[\vec{\ddot{a}} + 3\left(\frac{v}{c}\right)^2 \vec{a}(\vec{a}\cdot\vec{v})\right] \quad (40)$$

where a s the acceleration, $\vec{\dot{a}} = da/dt$ is the time rate of change of acceleration, v is the velocity, $m_o$ is the rest mass of an electron, and $\tau = 6.24 \times 10^{-24}$ sec. $\gamma = \left[1 - (v/c)^2\right]^{-1/2} \approx 3$ for 1 MeV electrons.

The equation of motion we need to solve is

$$e\vec{v} \times \vec{B} + \vec{F}_r = \gamma m_o \vec{a} \quad (41)$$

The solution to Eq. (41) for the velocity is:

$$v = v_o \exp\left[-\tau\omega_o^2 N\gamma\left(1+3\gamma^2\bar{\beta}^2\right)\right]\exp i\left[\left(\frac{\omega_o}{\gamma}\right)\left\{1+\left[\tau\omega_o N\gamma^2\left(1+3\gamma^2\bar{\beta}^2\right)\right]^2\right\}t\right] \quad (42)$$

for $N \ll 10^{14}$, and

$$v = v_o \exp\left[\frac{\left\{1-\left[2\tau\omega_o N\gamma^2\left(1+3\gamma^2\bar{\beta}^2\right)\right]^{1/2}\right\}t}{2\tau N\gamma^3\left(1+3\gamma^3\bar{\beta}^2\right)}\right]\exp i\left[\left(\frac{\omega_o^{1/2}}{\gamma}\right)\left\{1+\left[2\tau N\gamma^2\left(1+3\gamma^2\bar{\beta}^2\right)\right]^{-1/2}\right\}t\right] \quad (43)$$

for $N \gg 10^{14}$,

where $v_o$ is the initial velocity, $\bar{\beta}^2$ is the average value of $(v/c)^2$ over the time interval of interest (for the 1 MeV electrons we are concerned with, $\beta = 0.9$ initially), and $\omega_o = (eB/m_o)$ is the angular frequency of a nonrelativistic electron in a magnetic field of flux density $B = 0.6G = 0.6 \times 10^{-1}$ W/m$^2$ ($\omega_o = 1.05 \times 10^7$ /sec). In this notation, $v_x$ is the real part of $v$, and $v_y$ is the imaginary part of $v$.

Since the angular frequency, $\omega$, of the electron orbit (which is spiralling in toward the center of rotation) is approximately constant, we have:

$$\frac{a}{a_o} \approx \frac{\omega^2 R}{\omega^2 R_o} \approx \frac{R}{R_o} \quad (44)$$

$$\frac{a}{a_o} \approx \frac{\omega R}{\omega R_o} \approx \frac{v}{v_o} \quad , \quad (45)$$

where Eq. 42 is used in Eq. 45 for $N \ll 10^{14}$; and Eq. 43 is used in Eq. 45 for $N \gg 10^{14}$. In Eq. 43, the 1 can be neglected in the real exponential since it is small compared with the square root term for $N > \sim 10^{15}$ which is the case we are interested in.

In this case, the angular frequency is greatly reduced,

$$\omega = \left(\frac{\omega_o^{1/2}}{\gamma}\right)\left[2\tau N\gamma^2\left(1+3\gamma^2\bar{\beta}^2\right)\right]^{-1/2} < 6 \times 10^5 / \text{sec for } N > \sim 10^{15}. \quad (46)$$

This compares with an angular frequency for a single electron,

$$\omega = \left(\frac{eB}{\gamma m_o}\right) = \frac{1.05 \times 10^7 / \text{sec}}{3} = 3.50 \times 10^6 / \text{sec for } N = 1. \quad (47)$$

For N > ~ $10^{15}$, the radiation reaction force is obtained by applying Eq. 43 to Eq. 40:

$$F_r = Nm_o \tau \gamma^4 v \left(1 + 3\gamma^2 \beta^2\right) \left| \frac{2\omega_o}{v^2 \left(2\tau N v^2\right)\left(1 + 3\gamma^2 \overline{\beta}^2\right)} \right| \tag{48}$$

$$\approx m_o v \omega_o, \text{ for N > ~ } 10^{15}.$$

This is equal to the Lorentz force, $F_L$, acting on the electron by means of the geomagnetic field:

$$F_L = evB = v(eB) = v(m_o \omega_o) = F_r \text{ for N > ~ } 10^{15}. \tag{49}$$

For a radiation reaction damping force that can get as large as the Lorentz force, it may not be possible to have as many as N > ~ $10^{15}$ electrons radiating coherently.

Although the United States has conducted only one high-altitude, high-yield nuclear test, Starfish, the Soviet Union (and probably others) have conducted many such tests. EMP does not recognize national borders. Many bordering and nearby countries have been within the line-of-sight horizon of the Soviet EMP producing nuclear bursts. Yet, there have been no EMP complaints from these countries. There have been no reported power or communication outages caused by these tests.

In the event of an all-out nuclear attack, the United States might likely explode scores of our own nuclear weapons at high altitudes over our country as a defensive countermeasure. In this case, the EMP from our own weapons could unavoidably be a threat to the power grid. Although each nuclear burst would have a limited local effect, the combination of scores of sequential bursts may have nationwide impact.

**CONCLUSION**

The energy of the TEMP is one millionth of the bomb's energy release spread out over millions of square miles. Certainly, the lack of damage to both the power and the communications systems in Hawaii from the 1.4-Mt Starfish blast counters the view that the effects of EMP are devastating to such systems.

The telecommunications industry concluded, "Based upon results of testing done to date, the Task Force believes that significantly large portions of the Public Telecommunications Networks (PTNs) would survive a HEMP [high-altitude EMP] attack... " ([21], p. ES-1). One may expect the telecommunications system to be far more vulnerable than its rugged counterpart, the electric power system. The telecommunications system typically operates at ~ 10s of volts with a breakdown voltage ~ 100s of volts. In comparison, the electric power system looks extremely hardy, operating at ~ 100s of kilovolts, with a breakdown voltage ~ millions of volts. If significantly large portions of the telecommunications networks would survive an EMP attack, then one may expect an even more robust response from the electric power system. Even the electronic power system control equipment may be better protected than its telecommunications counterpart. Due to its compact nature, power system control equipment has little direct coupling with EMP. Indirect coupling of the TEMP with this equipment through high voltage power lines involves large attenuation of the pulse in both entering and going through transformers as established in both the Sandia and Swedish studies.

Based upon the analyses presented in this paper and in Refs. [7-8], it appears highly improbable if not impossible that the EMP from a single nuclear burst could blackout this nation's power grid. It would be practically impossible for the EMP to cause widespread damage to the U.S. transmission line system. With the exception of isolated cases, it appears highly unlikely that EMP could produce extensive damage to the U.S. distribution grid. A single nuclear device exploded at high altitude will not render vital electrical services inoperable across the entire United States as has been suggested in many media references.

Concurrent multiple bomb bursts will **not** have an additive TEMP effect, and will even interfere to produce less TEMP than a single burst.

## ACKNOWLEDGMENTS

I want to express my appreciation to Felipe Garcia, Gene Salamin, Charles Vittitoe, and William Hassenzahl for their helpful comments and criticisms. I especially wish to express sincere gratitude to John Dougherty for his continued interest and support of this work. Special thanks are due my wife, Laverne, for her encouragement and for her patience during the writing of this paper.